\documentclass[aps,prb,twocolumn,superscriptaddress,floatfix,showpacs,10pt,letterpaper]{revtex4-1}

\pdfoutput=1

\begin{document}

\begin{titlepage}

\title{Complete classification of 1D gapped quantum phases\\
 in interacting spin systems}

\author{Xie Chen}
\affiliation{Department of Physics, Massachusetts Institute of
Technology, Cambridge, Massachusetts 02139, USA}
\author{Zheng-Cheng Gu}
\affiliation{Kavli Institute for Theoretical Physics, University of California, Santa Barbara, CA 93106, USA}
\author{Xiao-Gang Wen}
\affiliation{Department of Physics, Massachusetts Institute of
Technology, Cambridge, Massachusetts 02139, USA}

\begin{abstract}

Quantum phases with different orders exist with or without breaking the
symmetry of the system. Recently, a classification of gapped quantum phases
which do not break time reversal, parity or on-site unitary symmetry has been
given for 1D spin systems in  [X. Chen, Z.-C. Gu, and X.-G. Wen, Phys.  Rev. B
\textbf{83}, 035107 (2011); arXiv:1008.3745]. It was found that, such symmetry
protected topological (SPT) phases are labeled by the projective
representations of the symmetry group which can be viewed as a symmetry
fractionalization. In this paper, we extend the classification of 1D gapped
phases by considering SPT phases with combined time reversal, parity, and/or
on-site unitary symmetries and also the possibility of symmetry
breaking. We clarify how symmetry fractionalizes with combined symmetries 
and also how symmetry fractionalization coexists with symmetry breaking.
 In this way, we obtain a complete classification of gapped quantum
phases in 1D spin systems. We find that in general, symmetry fractionalization,
symmetry breaking and  long range entanglement(present in 2 or higher
dimensions) represent three main mechanisms to generate a very rich set of
gapped quantum phases.  As an application of our classification, we study the
possible SPT phases in 1D fermionic systems, which can be mapped to spin
systems by Jordan-Wigner transformation.

\end{abstract}


\maketitle

\vspace{2mm}

\end{titlepage}

\section{Introduction}

Quantum phases of matter with exotic types of order have continued to emerge
over the past decades. Examples include
fractional quantum Hall states\cite{TSG8259,L8395}, 1D Haldane
phase\cite{H8364}, chiral spin liquids,\cite{KL8795,WWZ8913} $Z_2$ spin
liquids,\cite{RS9173,W9164,MS0181} non-Abelian fractional quantum Hall
states,\cite{MR9162,W9102,WES8776,RMM0899} quantum orders characterized by
projective symmetry group (PSG),\cite{W0213} topological
insulators\cite{KM0501,BZ0602,KM0502,MB0706,FKM0703,QHZ0824}, etc. Why are
there different orders? What is a general framework to understand all these
seemingly very different phases? How to classify all possible phases and
identify new ones? Much effort has been devoted to these questions, yet the
picture is not complete.

First, we want to emphasize that quantum phase is a property of a class of
Hamiltonians, not of a single Hamiltonian. We call such a class of Hamiltonian
an H-class.  For an H-class, of a certain dimension and with possible symmetry
constraints, we ask whether the Hamiltonians in it are separated into different
groups by phase transition and hence form different phases. Two Hamiltonians in
an H-class are in the same/different phase if they can/cannot be connected
\emph{within the H-class} without going through phase transition. We see that
without identifying the class of Hamiltonians under consideration, it is not
meaningful to ask which phase a Hamiltonian belongs to. Two Hamiltonians can
belong to the same/different phases if we embed them in different H-classes.

For an H-class with certain symmetry constraints, one mechanism leading to
distinct phases is symmetry breaking.\cite{L3726,GL5064} Starting from
Hamiltonians with the same symmetry, the ground states of them can have
different symmetries, hence resulting in different phases. This symmetry
breaking mechanism for phases and phase transitions
is well understood.\cite{L3726,LanL58} 

However, it has been realized that systems can be in different phases even
without breaking any symmetry. Such phases are often said to be `topological'
or `exotic'.  However, the term `topological' in literature actually refers to
two different types of quantum order.

The first type has `intrinsic' topological order. This type of order is defined
for the class of systems without any symmetry constraint, which corresponds to the
original definition of `topological order'.\cite{W8987,W9039} That is, it refers
to quantum phases in an H-class which includes all local Hamiltonians (of a
certain dimension).  If we believed that Landau symmetry breaking theory describes
all possibles phases, this whole H-class would belong to the same phase as there
is no symmetry to break. However, in two and three dimensions, there are
actually distinct phases even in the H-class that has no symmetries. These
phases have universal properties stable against any small local perturbation to
the Hamiltonian. To change these universal properties, the system has to go
through a phase transition. Phases in this class include quantum Hall
systems\cite{WN9077}, chiral spin liquids,\cite{KL8795,WWZ8913} $Z_2$ spin
liquids,\cite{RS9173,W9164,MS0181} quantum double model\cite{K0302} and
string-net model\cite{LW0510}. Ground states of these systems have `long-range
entanglement' as discussed in \Ref{CGW1038}. 

The `topological' quantum order of the second type is `symmetry protected'. The class of
systems under consideration have certain symmetry and the ground states have
only short-range entanglement\cite{CGW1038}, like in the symmetry breaking
case. However unlike in symmetry breaking phases, the ground states have the same
symmetry as the Hamiltonian and, even so, the ground states can be in different
phases. This quantum order is protected by symmetry; as according to the discussion
in \Ref{CGW1038}, if the symmetry constraint on the class of systems is
removed, all short-range entangled states belong to the same phase. Only when
symmetry is enforced, can short range entangled states with the same symmetry
belong to different phases.  Examples of this type include the Haldane
phase\cite{H8364} and topological
insulators.\cite{KM0501,BZ0602,KM0502,MB0706,FKM0703,QHZ0824}

Phases in these two classes share some similarities. For example, they both are
beyond Landau symmetry breaking theory. Also quantum Hall systems and
topological insulators both have stable gapless edge
states.\cite{W9125,BHT0657,FK0702} However, the latter requires symmetry
protection while the former does not.

Despite the similarities, these two classes of topological phases are
fundamentally different, as we can see from quantities that are sensitive to
long-rang entanglement.  For example, `intrinsic' topological order has a
robust ground state degeneracy that depends on the topology of the
space.\cite{W8987,W9039} The ground states with `intrinsic' topological order
also have non-zero topological entanglement entropy\cite{LW0605,KP0604} while
ground states in `symmetry protected' topological phases are short range
entangled and therefore have zeros topological entanglement entropy. Also,
the low energy excitations in `symmetry protected' topological phases do not
have non-trivial anyon statistics, unlike in `intrinsic' topological phases.
\cite{ASW8422,WWZ8913,K0602}

In the following discussion, we will use `topological phase' to refer only to the
first type of phases (\ie `intrinsic' topological phases). For the second type,
we will call them `Symmetry Protected Topological' (SPT) phases, as in
\Ref{GW0931}. Similar to the quantum orders characterized by PSG,\cite{W0213} different
SPT phases are also characterized by the projective representations of the
symmetry group of the Hamiltonian.\cite{CGW1107}
The PSG and projective
representations of a symmetry group can be viewed as a `fractionalization' of
the symmetry.  Thus, we may say that different SPT phases are caused by
`symmetry fractionalization'.

\begin{figure}
\begin{center}
\includegraphics[scale=0.4]{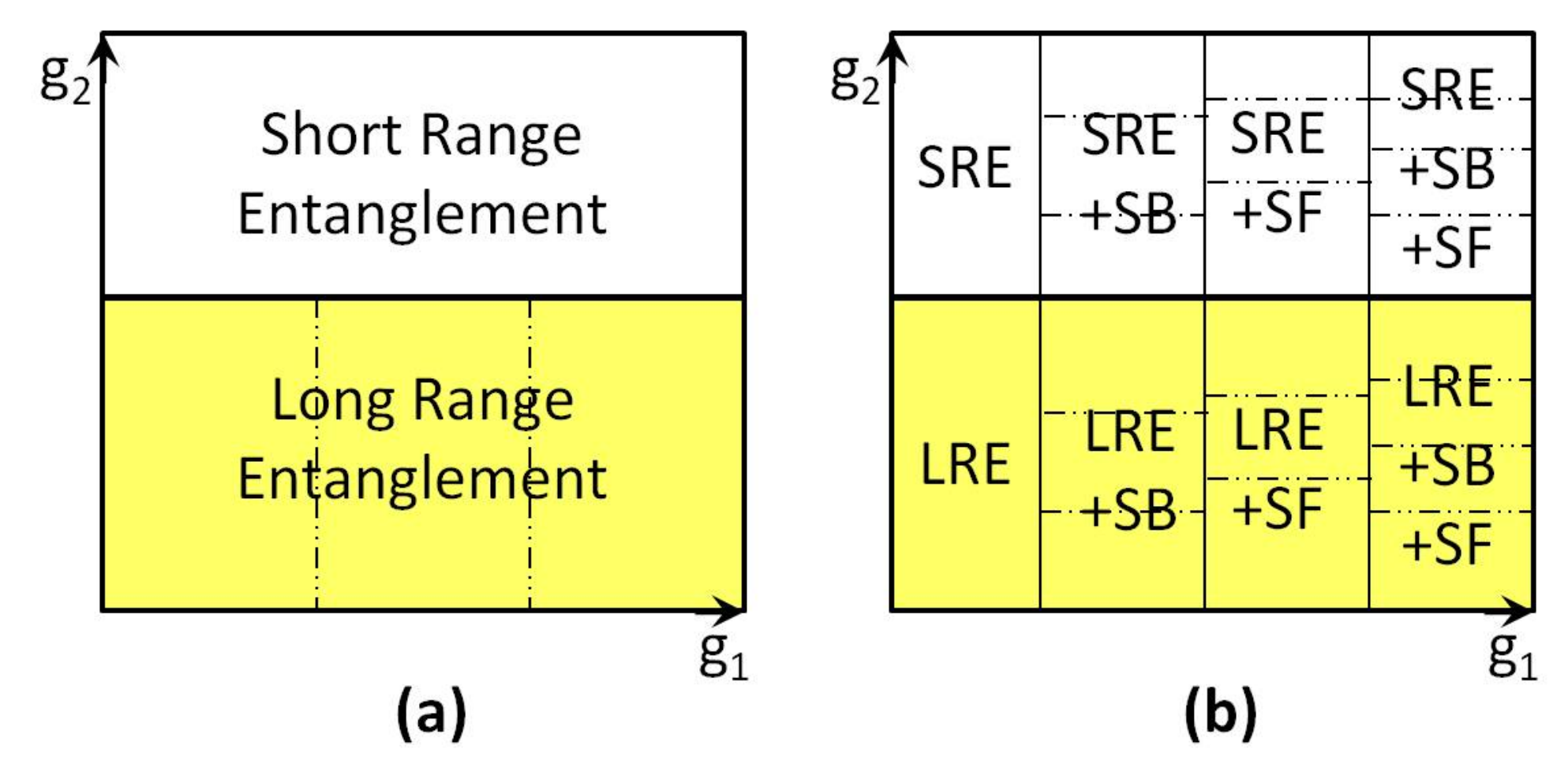}
\end{center}
\caption{
(Color online) (a) The possible phases for class of Hamiltonians without any
symmetry.  (b) The possible phases for class of Hamiltonians with some
symmetries.  Each phase is labeled by the phase separating mechanisms involved.
The shaded regions in (a) and (b) represent the phases with long range
entanglement. SRE stands for short range entanglement, LRE for long range
entanglement, SB for symmetry breaking, SF for symmetry fractionalization.
}
\label{fig:phase}
\end{figure}

Long-rang entanglement, symmetry fractionalization, and symmetry breaking
represent three different mechanisms to separate phases and can be combined to
generate a very rich quantum phase diagram. Fig. \ref{fig:phase} shows a phase
diagram with possible phases generated by these three mechanisms. In order to
identify the kind of quantum order in a system, we first need to know whether
topological orders exist. That is, whether the ground state has long range
entanglement.  Next, we need to identify the symmetry of the system (of the
Hamiltonian and allowed perturbation). Then we can find out whether all or part
of the symmetry is spontaneously broken in the ground state. If only part is broken, what is the SPT order due to
the fractionalization of the unbroken symmetry. Combining these data together
gives a general description of a quantum phase. Most of the phases studied
before involve only one of the three mechanisms. Examples where two of
them coexist can be found in \Ref{WWZ8913,KLW0902} which combine long
range entanglement (the intrinsic topological order) and symmetry breaking and
in \Ref{W0213,KLW0834,KW0906,YFQ1070} which combine long range entanglement
and symmetry fractionalization. In fact, the PSG provides a quite
comprehensive framework for symmetry fractionalization in topologically ordered
states.\cite{W0213,KLW0834,KW0906}

Based on this general understanding of quantum phases, we address the following
question in this paper: what quantum phases exist in one-dimensional gapped
spin systems. The systems we consider can have any finite strength finite range
interactions among the spins. 

In \Ref{CGW1107}, we gave a partial answer to this question. We first showed
that one-dimensional gapped spin systems do not have non-trivial topological
order. So to understand possible 1D gapped phases, we just need to
understand symmetry fractionalization and symmetry breaking in short-range
entangled  states.  In other words, quantum phases are only different because
of symmetry breaking and symmetry fractionalization. 

In  \Ref{CGW1107}, we
then considered symmetry fractionalization, and gave a classification of
possible SPT phases with time reversal, parity and on-site unitary symmetry
respectively. In this paper, we complete the classification by considering SPT
phases of combined time reversal, parity and/or on-site unitary symmetry and
finally incorporate the possibility of symmetry breaking. We find that in 1D
gapped spin system with on-site symmetry of group $G$, the quantum phase are
basically labeled by:\\
 (1) the unbroken symmetry subgroup $G'$\\
 (2) projective representation of the unbroken part of on-site unitary and anti-unitary symmetry respectively\\
 (3) `projective' commutation relation between representations of 
      unbroken symmetries.\\
Here the projective representation and the `projective' commutation relation
represent the symmetry fractionalization. Parity is not an on-site symmetry and its SPT phases are not characterized by projective representations. The classification involving parity does not fall into the general framework above, but proceeds in a very similar way, as we will show in section \ref{CS}. Actually, (2) and (3) combined gives the projective representation of $G'$ and if parity is present, it should be treated as an anti-unitary $Z_2$ element. Our result is consistent with those obtained by Schuch et al\cite{SPC1032}.

Our discussion is based on the matrix product state
representation\cite{FNW9243, PVW0701} of gapped 1D ground states. The matrix
product formalism allows us to directly deal with interacting systems and its
entangled ground state. In particular, SPT order of a system can be identified directly from the way matrices in the representation transform under symmetries.\cite{PTB1039}

Moreover, symmetry breaking in entangled systems can be represented in a
nice way using matrix product states.\cite{FNW9243,PVW0701,N9665} The traditional understanding of symmetry
breaking in quantum systems actually comes from intuition about classical
systems. For example, in the ferromagnetic phase of classical Ising model, the
spins have to choose between two possible states: either all pointing up or all
pointing down. Both states break the spin flip symmetry of the system.
However, for quantum system, the definition of symmetry breaking becomes a
little tricky. In ferromagnetic phase of quantum Ising model, the ground space
is two fold degenerate with basis states $|\uparrow \uparrow ...\uparrow\>$,
$|\downarrow \downarrow ...\downarrow\>$. Each basis state breaks the spin flip
symmetry of the system. However, quantum systems can exist in any superposition
of the basis states and in fact the superposition $\frac{1}{\sqrt{2}}|\uparrow
\uparrow ...\uparrow\> + \frac{1}{\sqrt{2}}|\downarrow \downarrow
...\downarrow\>$ is symmetric under spin flip. What is meant when a
\emph{quantum} system is said to be in a symmetry breaking phase?
Can we understand symmetry breaking in a quantum system
without relying on its classical picture?

In matrix product representation, the symmetry breaking pattern can be seen
directly from the matrices representing the ground state. If we choose the symmetric ground state in the
ground space and write it in matrix product form, the matrices can be reduced to a
block diagonal `canonical form'. The `canonical form' contains more than one
block if the system is in a symmetry breaking phase. If symmetry is not
broken, it contains only one block.\cite{FNW9243,PVW0701,N9665} Hence, the
`canonical form' of the matrices give a nice illustration of the symmetry
breaking pattern of the system. This relation will be discussed in more detail
in section \ref{SB}.

Our classification is focused on 1D interacting spin systems, however it also applies to 1D
interacting fermion systems as they are related by Jordan-Wigner
transformation. As an application of our classification result, we study
quantum phases(especially SPT phases) in gapped 1D fermion systems. Our
result is consistent with previous studies.\cite{TPB1102,LK1103} 

The paper is organized as follows: in section \ref{SPT}, we review the previous
classification results of SPT phases with time reversal, parity and on-site
unitary symmetry respectively. We also introduce notations for matrix product
representation; in section \ref{CS}, we present classification result of SPT
phases with combined time reversal, parity and/or on-site unitary symmetry; in
section \ref{SB}, we incorporate the possibility of symmetry breaking; in
section \ref{fermion}, we apply classification results of spins to the study of phases in 1D fermion
systems; and finally we conclude in section \ref{conclude}.

\section{Review: Matrix product states and SPT classification}
\label{SPT}

In \Ref{CGW1107}, we considered the classification of SPT phases with time
reversal, parity, and on-site unitary symmetry respectively. Instead of
starting from Hamiltonians, we classified 1D gapped ground states which do not
break the symmetry of the system. The set of states under consideration can be
represented as short-range correlated matrix product states and we used the
local unitary equivalence between gapped ground states, which was established
in \Ref{CGW1038}, to classify phases. Here we introduce the matrix product
representation and give a brief review of previous classification result and
how it was achieved.

Matrix product states give an efficient representation of 1D gapped spin states\cite{VC0623,SWV0804} and hence provide a useful tool to deal with strongly interacting systems with many-body entangled ground states. A matrix product states (MPS) is expressed as
\begin{equation}
\label{MPS}
|\phi\rangle = \sum_{i_1,i_2,...,i_N} 
\Tr (A_{i_1}A_{i_2}...A_{i_N})|i_1i_2...i_N\rangle
\end{equation}
where $i_k=1...d$ with $d$ being the physical dimension of a spin at each site, $A_{i_k}$'s are $D\times D$ matrices on site $k$ with $D$ being the inner dimension of the MPS. In our previous studies\cite{CGW1107} and also in this paper, we consider states which can be represented with a finite inner dimension $D$ and assume that they represent all possible phases in 1D gapped systems. In our following discussion, we will focus on states represented with site-independent matrices $A_{i}$ and discuss classification of phases with/without translational symmetry. Non-translational invariant systems have in general ground states represented by site dependent matrices. However, site dependence of matrices does not lead to extra features in the phase classification and their discussion involves complicated notation. Therefore, we will not present the analysis based on site dependent MPS. A detailed discussion of site dependent MPS can be found in \Ref{CGW1107} and all results in this paper can be obtained using similar method.

A mathematical construction that will be useful is the double tensor
\begin{equation}
\mathbb{E}_{\alpha\gamma,\beta\chi}=\sum_{i}
A_{i,\alpha\beta} \times (A_{i,\gamma\chi})^*
\end{equation} 
of the MPS. $\mathbb{E}$ is useful because it uniquely determines the a matrix product state up to a local change of basis on each site\cite{NC2000,PVW0701}. Therefore, all correlation and entanglement information of the state is contained in $\mathbb{E}$ and can be extracted.

First we identify the set of matrix product states that need to be considered for the classification of SPT phases. The ground state of SPT phases does not break any symmetry of the system and hence is non-degenerate. The unique ground state must be short-range correlated due to the existence of gap, which requires that $\mathbb{E}$ has a non-degenerate largest eigenvalue(set to be $1$)\cite{FNW9243, PVW0701}. This is equivalent to an `injectivity' condition on the matrices $A_{i}$. That is, for large enough $n$, the set of matrices corresponding to physical states on $n$ consecutive sites $\mathcal{A}_I = A_{i_1}...A_{i_n}$($I \equiv i_1...i_n$) span the space of $D \times D$ matrices\cite{PVW0701}. On the other hand, all MPS satisfying this injectivity condition are the unique gapped ground states of a local Hamiltonian which has the same symmetry\cite{FNW9243, PVW0701}. Therefore, we only need to consider states in this set.

The symmetry of the system and hence of the ground state sets a non-trivial transformation condition on the matrices $A_{i}$. With on-site unitary symmetry of group $G$, for example, the $A_{i}$'s transform as\cite{PWS0802,PTB1039}
\begin{equation}
\sum_{j} u(g)_{ij} A_{j} = \alpha(g)R^{-1}(g)A_{i}R(g)
\label{eqn:U}
\end{equation}
where $u(g)$ is a linear representation of $G$ on the physical space, $\alpha(g)$ is a one-dimensional representation of $G$. One important realization from this equation is that, in order to satisfy this equation, $R(g)$ only has to satisfy the multiplication rule of group $G$ up to a phase factor\cite{PTB1039}. That is, $R(g_1g_2)=\omega(g_1,g_2)R(g_1)R(g_2)$, $|\omega(g_1,g_2)|=1$. $\omega(g_1,g_2)$ is called the factor system. $R(g)$ is hence a projective representation of group $G$ and belongs to different equivalence classes labeled by the elements in the second cohomology group of $G$, $\{\omega|\omega \in H^2(G,\mathbb{C})\}$. 

We showed in \Ref{CGW1107} that two matrix product states symmetric under $G$ are in the same SPT phase if and only if they are related to $R(g)$ in the same equivalence class $\omega$. For two states with equivalent $R(g)$, we constructed explicitly a smooth path connecting the Hamiltonian for the first state to that for the second state without closing the gap or breaking the symmetry of the system. In this way, we gave a `local unitary transformation' as defined in \Ref{CGW1038} connecting the two state and showed that they are in the same SPT phase. On the other hand, for two states associated with inequivalent $R(g)$, we showed that no `local unitary transformation' could connect them without breaking the symmetry. Therefore, they belong to different SPT phases. 

If translation symmetry is required in addition to symmetry $G$, $\alpha(g)$ is also a good quantum number and cannot be changed without breaking translation symmetry. Therefore, SPT phases with on-site unitary symmetry and translation symmetry are labeled by $\{\alpha(g),\omega\}$, where $\alpha(g)$ is a one-dimensional representation of $G$ and $\omega$ is an element in $H^2(G,\mathbb{C})$.

\begin{figure}
\begin{center}
\includegraphics[width=3.2in]{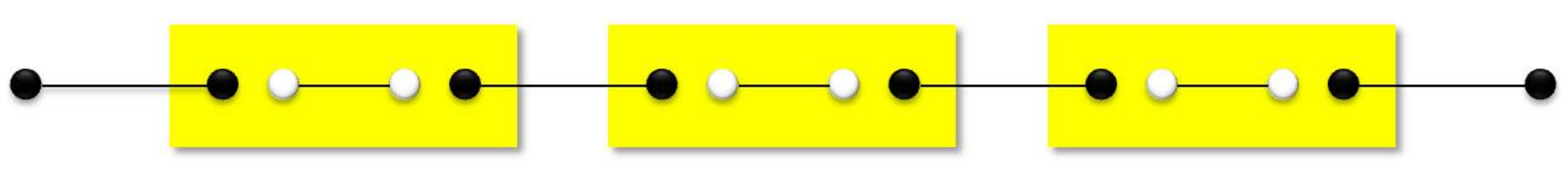}
\end{center}
\caption{
(Color online)
Representative states for different SPT phases. Each box represents one site, containing four spins. Every two connected spins form an entangled pair.}
\label{fig:rep_1}
\end{figure}

The projective representation can be interpreted in terms of boundary spins. A representative state in the phase labeled by $\{\alpha(g),\omega\}$ can be given as in Fig.\ref{fig:rep_1}. Each box represents one site, containing four spins. The symmetry transformations on the two black spins form projective representations of $G$, belonging to class $\omega$, $\omega'$ respectively. The factor systems of the two classes are related by $\omega(g_1,g_2)\times \omega'(g_1,g_2)=1$, that is $\omega'=\omega^*$. Therefore, the inter-site black pair can form a singlet under symmetry $G$. Suppose that the pair forms a 1D representation $\alpha_1(g)$ of $G$. The on-site white pair also forms a 1D representation $\alpha_2(g)$ of $G$. $\alpha_1(g)\alpha_2(g)=\alpha(g)$. It can be checked that, if written in matrix product representation, the matrices satisfy condition \ref{eqn:U}. Now look at any finite segment of the chain. There are unpaired black spins at each end of the chain, transforming under $G$ as projective representation $\omega$ and $\omega^*$. Different $\omega$ cannot be smoothly mapped to each other if the on-site linear symmetry is maintained. Therefore, by looking at the boundary of a finite chain, we can distinguish different SPT phases with on-site unitary symmetry.

For projective representations in the same class, we can always choose the phases of $R(g)$ such that $\omega(g_1,g_2)$ is the same. In the following discussion we will always assume that $\omega(g_1,g_2)$ is fixed for each class and the phase of $R(g)$ is chosen accordingly. But this does not fix the phase of $R(g)$ completely. For any 1D linear representation $\tilde{\alpha}(g)$, $\tilde{\alpha}(g)R(g)$ always has the same factor system $\omega(g_1,g_2)$ as $R(g)$. This fact will be useful in our discussion of the next section.

The classifications for SPT phases with time reversal and parity symmetry proceed in a similar way.

For time reversal, the physical symmetry operation is $T=v\otimes v... \otimes vK$, where $K$ is complex conjugation and $v$ is an on-site unitary operation satisfying $vv^*=I$, that is $T^2=I$ on each site. (We showed in \Ref{CGW1107} that if $T^2=-I$ on each site, there are no gapped symmetric phases with translation symmetry. Without translation symmetry, it is equivalent to the $T^2=I$ case.) The symmetry transformation of matrices $A_{i}$ is
\begin{equation}
\sum_{j} v_{ij} A^*_{j} = M^{-1} A_{i} M
\label{eqn:TR}
\end{equation}
where $M$ satisfies $MM^*=\beta(T) I = \pm I$. It can be shown, in a way similar to the on-site unitary case, that states with the same $\beta(T)$ can be connected with local unitary transformations that do not break time reversal symmetry while states with different $\beta(T)$ cannot. Therefore, $\beta(T)$ labels the two SPT phases for time reversal symmetry. We can again understand this result using boundary spins. Time reversal on the boundary spin can be defined as $\hat{T}=MK$. It squares to $\pm I$ depending on $\beta(T)$. Therefore, while time reversal acting on the physical spin at each site always squares to $I$, it can act in two different ways on the boundary spin, hence distinguishing two phases. $\hat{T}=MK$ forms a projective representation of time reversal on the boundary spin, with $\hat{T}^2=\pm I$. The result is unchanged if translational symmetry is required. 

For parity symmetry, the physical symmetry operation is $P=w\otimes w...\otimes w P_1$, where $P_1$ is exchange of sites and $w$ is on-site unitary satisfying $w^2=1$. As parity symmetry cannot be established in disordered systems, we will always assume translation invariance when discussing parity. The symmetry transformation of matrices $A_{i}$ is
\begin{equation}
\sum_{j} w_{ij} A^T_{j} = \alpha(P) N^{-1} A_{i} N
\label{eqn:P}
\end{equation}
where $\alpha(P)=\pm 1$ labels parity even/odd and $N^T=\beta(P)N=\pm N$. The four SPT phases are labeled by $\{\alpha(P),\beta(P)\}$. A representative state can again be constructed as in Fig.\ref{fig:rep_1}. The black spins form an entangled pair $(N\otimes I)\sum_i|i\>\otimes |i\>$, $i=1...D$, $D$ being the dimension of $N$. ($|i\>\otimes |j\>$ denotes a product state of two spins in state $|i\>$ and $|j\>$ respectively. This is equivalent to notation $|i\>|j\>$ and $|ij\>$ in different literatures.) The white spins form an entangled pair $|0\>\otimes |1\> + \alpha(P)\beta(P)|1\>\otimes |0\>$. Parity is defined as reflection of the whole chain. It can be checked that if written as matrix product states, the matrices satisfy condition Eq.\ref{eqn:P} and parity of the black pair is determined by $\beta(P)$. Therefore, $\beta(P)$ can be interpreted as even/oddness of parity between sites and $\alpha(P)$ represents total even/oddness of parity of the whole chain.

\section{Classification with combined symmetry}
\label{CS}

In this section we are going to consider the classification of SPT phases with
combined translation, on-site unitary, time reversal, and/or parity symmetry in
1D gapped spin systems.  The ground state does not break any of the combined
symmetry and can be described as a short-range correlated matrix product state.
For each combination of symmetries, we are going to list  all possible SPT
phases and give a label and representative state for each of them. In the end of 
this section, we comment on the general scheme to classify SPT phases with all possible kinds of symmetries in 1D gapped spin systems.

\subsection{Parity + On-site $G$}
Consider a system symmetric under both parity $P=w\otimes w...\otimes w P_1$
and on-site unitary of group $G$ $u(g)\otimes u(g) ... \otimes u(g)$.
Eq.\ref{eqn:U} and \ref{eqn:P} give transformation rules of matrices $A_{i}$
under the two symmetries separately in terms of $\alpha(g)$, $R(g)$,
$\alpha(P)$, $N$.  $\alpha(g)$ labels the 1D representation the state forms under $G$, $R(g)\in \omega$ is the projective representation of $G$ on the boundary spin, $\alpha(P)$ labels parity even or odd and $N^T=\beta(P)N=\pm N$ corresponds to parity even/odd between sites.

Moreover, parity and on-site $u(g)$ commute. First, it is easy to see that $P_1$ and $u(g) \otimes u(g)...\otimes u(g)$ commute. Therefore, $P_2=w \otimes w ... \otimes w$ must also commute with $u(g) \otimes u(g)...\otimes u(g)$. WLOG, we will consider the case where $w$ and $u(g)$ commute, $wu(g)=u(g)w$. This leads to certain commutation relation between $N$ and $R(g)$ as shown below.

If we act parity first and on-site symmetry next, the matrices are transformed as
\begin{equation}
\begin{array}{lll}
A_i   \xrightarrow{P} A'_i &= & \sum_j w_{ij} A^T_j =\alpha(P)N^{-1}A_iN \\
A'_i  \xrightarrow{G} A''_i &= & \sum_j u_{ij}(g) A'_j \\
                                        &=& \alpha(P)N^{-1}\left(\sum_j u_{ij}A_j\right)N \\
                                        &= & \alpha(g)\alpha(P) N^{-1}R^{-1}(g)A_iR(g)N
\end{array}
\end{equation}
Combining the two steps together we find that 
\begin{equation}
\sum_{j} \sum_{k} u_{ij}(g)  w_{jk} A^T_k = \alpha(g)\alpha(P) N^{-1}R^{-1}(g)A_iR(g)N
\label{eqn:PG}
\end{equation}
If on-site symmetry is acted first and then parity follows, the matrices are transformed as
\begin{equation}
\begin{array}{lll}
A_i  \xrightarrow{G} A'_i &=& \sum_j u_{ij}(g) A_j= \alpha(g)R^{-1}(g)A_iR(g) \\
A'_i  \xrightarrow{P} A''_i &=& \sum_j w_{ij} (A')^T_j \\
                                        &=& \alpha(g)R^T(g)(\sum_j w_{ij} A^T_j)(R^T)^{-1}(g) \\
                                        &=& \alpha(P)\alpha(g) R^T(g)N^{-1}A_iNR^*(g)
\end{array}      
\end{equation}
The combined operation is then
\begin{equation}
\sum_{j} \sum_k w_{ij}  u_{jk}(g) A^T_k 
= \alpha(P)\alpha(g) R^T(g)N^{-1}A_iNR^*(g)
\label{eqn:GP}
\end{equation}
Because $w$ and $u(g)$ commute, $\sum_{j} u_{ij}(g) w_{jk}=\sum_{j} w_{ij} u_{jk}(g)$. Therefore, the combined operation in Eq. \ref{eqn:PG} and \ref{eqn:GP} should be equivalent.
\begin{equation}
N^{-1}R^{-1}(g)A_iR(g)N = R^T(g)N^{-1}A_iNR^*(g)
\end{equation}
This condition is derived for matrices on each site, $i=1...d$. However, it is easy to verify that it also holds if $n$ consecutive sites are combined together with representing matrices $\mathcal{A}_I=A_{i_1}A_{i_2}...A_{i_n}$.
\begin{equation}
N^{-1}R^{-1}(g)\mathcal{A}_IR(g)N = R^T(g)N^{-1}\mathcal{A}_INR^*(g)
\end{equation}
As $\mathcal{A}_I$ is injective(spans the whole space of $D\times D$ matrices), we find $R(g)NR^T(g)N^{-1} \propto I$. That is
\begin{equation}
N^{-1}R(g)N=e^{i\theta(g)}(R^T)^{-1}(g)=e^{i\theta(g)}R^*(g)
\label{eqn:R_gamma}
\end{equation}
Different $e^{i\theta(g)}$ corresponds to different `projective' commutation relations between parity and on-site unitary. It must satisfy certain conditions. As
\begin{equation}
\begin{array}{lll}
N^{-1}R(g_1g_2)N & = & e^{i\theta(g_1g_2)}R^*(g_1g_2)\\
                              & = & e^{i\theta(g_1g_2)}\omega^{-1}(g_1,g_2)R^*(g_1)R^*(g_2) \\
N^{-1}R(g_1g_2)N & = & \omega(g_1,g_2)N^{-1}R(g_1)NN^{-1}R(g_2)N \\
                               & = & \omega(g_1,g_2)e^{i\theta(g_1)}R^*(g_1)e^{i\theta(g_2)}R^*(g_2)
\end{array}
\end{equation}
Therefore
\begin{equation}
e^{i\theta(g_1g_2)}e^{-i\theta(g_1)}e^{-i\theta(g_2)}=\omega^2(g_1,g_2)
\end{equation}
Hence, $\omega^2$ must be trivial. WLOG, assume that the factor systems we have chosen(as discussed in section \ref{SPT}) satisfy $\omega^2=1$ and $e^{i\theta(g)}$ forms a linear representation of $G$, denoted by $\gamma(g)$.

Let us interpret this result.

First we see that the combination of parity with on-site $G$ restricts the projective representation that can be realized on the boundary spin to those $\omega$ that square to identity. This can be clearly seen from the structure of the representative state as in Fig.\ref{fig:rep_1}. Because of on-site symmetry $G$, the left and right black spin in a pair form projective representation in class $\omega$ and $\omega^*$ respectively. If the chain has further reflection symmetry, $\omega^*=\omega$ and therefore $\omega^2=1$. However, if $\omega^2\neq 1$, then $\omega^* \neq \omega$. The chain has a direction and cannot have reflection symmetry.

Different $\gamma(g)$ corresponds to different `projective' commutation relation between parity and on-site $G$ on the boundary spin. For example, suppose $G=Z_2$. Consider the state as in Fig.\ref{fig:rep_1} where each black pair consists of two qubits. Suppose that parity on the black pair is defined as exchanged of the two qubits($P_1$) and unitary operation $P_2=Z\otimes Z$ on the two qubits. $Z_2$ symmetry on the pair can be defined as $Z\otimes Z$ or $X\otimes X$. They both commute with parity. However, if we only look at one end of pair, $Z_2$ either commute or anti-commute with $P_2$. Hence these two cases correspond to two different $\gamma(Z_2)$. Because of this, the two phases cannot be connected without breaking the symmetry group generated by parity and $Z_2$.

However, $\gamma(g)$ can be changed by changing the phase of $R(g)$. Remember that the phase of $R(g)$ is only determined(by fixing $\omega$) up to a 1D representation, $\tilde{\alpha}(g)$. From Eq.\ref{eqn:R_gamma}, we can see that if the phase of $R(g)$ is changed by $\tilde{\alpha}(g)$, $\gamma(g)$ is changed to $\gamma(g)/\tilde{\alpha}^2(g)$. Therefore, $\gamma(g)$ and $\gamma'(g)$ which differ by the square of another 1D representation $\tilde{\alpha}(g)$ are equivalent. On the other hand, for a fixed $\omega$($\omega^2=1$) any $\gamma(g)$ can be realized as the `projective' commutation relation, as we will show in Appendix \ref{append_B}.

Therefore, with commuting parity and on-site unitary symmetry, SPT phases in 1D spin chain can be classified by the following data: 
\begin{enumerate}
\item $\alpha(P)$, parity even/odd; 
\item $\beta(P)$, parity even/odd between sites; 
\item $\alpha(g)$, 1D representation of $G$; 
\item $\omega$, projective representation of $G$ on boundary spin, $\omega^2=1$;
\item $\gamma(g) \in \mathcal{G}/\mathcal{G}_2$, 1D representation of $G$ related to commutation relation between parity and on-site $G$, where $\mathcal{G}$ is the group of 1D representation of $G$, $\mathcal{G}_2$ is the group of 1D representation squared of $G$. 
\end{enumerate}

Following the method used in Appendix G of \Ref{CGW1107}, we can show that
states symmetric under parity and on-site $G$ are in the same SPT phase if and
only if they are labeled by the same set of data as given above. We will not
repeat the proof here. 

\begin{figure}
\begin{center}
\includegraphics[width=3.2in]{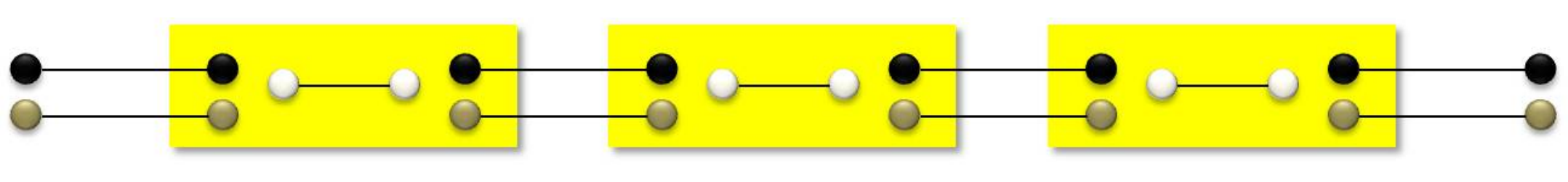}
\end{center}
\caption{
(Color online)
Representative states for different SPT phases. Each box represents one site, containing six spins. Every two connected spins form an entangled pair.}
\label{fig:rep_2}
\end{figure}

A representative state for each phase labeled by $\alpha(P)$, $\beta(P)$,
$\alpha(g)$, $\omega$ and $\gamma(g)$ can be constructed as in
Fig.\ref{fig:rep_2}. We will describe the state of each pair and how it
transforms under symmetry operations. The parameters describing the state will
then be related to the phase labels.
 
Each entangled pair is invariant under $P$ and on-site $G$. First, the on-site
white pair forms 1D representations $\eta(P)$ and $\eta(g)$ of parity and $G$.
For projective representation class $\omega$ that satisfies $\omega^2=1$ and
any 1D representation $\lambda(g)$, we show in appendix \ref{append_A} and
\ref{append_B} that there always exist projective representation $R(g) \in
\omega$ and symmetric matrix $N$($N^T=N$) such that
$N^{-1}R(g)N=\lambda(g)R^*(g)$. Suppose that $R(g)$ and $N$ are $D$-dimensional
matrices, choose the inter-site black pair to be composed of two
$D$-dimensional spins. Define parity on this pair to be exchange of two spins
and define on-site symmetry to be $R(g)\otimes R(g)$. If the state of the black
pair is chosen to be $(N\otimes I) \sum_i|i\rangle\otimes |i\rangle$, $i=1,...,D$, it
is easy to check that it is parity even, forms 1D representation $\lambda(g)$
for on-site $G$  and contains projective representation $\omega$ at each end.
Finally, define the state of the inter-site grey pair to be
$|0\>\otimes |1\>+\rho(P)|1\>\otimes |0\>$, $\rho(P)=\pm 1$. Parity acts on it as exchange of
spins. On-site $G$ acts trivially on it. The 1D spin state constructed in this
way is symmetric under parity and on-site unitary $G$ and belongs to the SPT
phase labeled by $\alpha(P)=\eta(P)\rho(P)$, $\beta(P)=\rho(P)$,
$\alpha(g)=\eta(g)\lambda(g)$, $\omega$, and $\gamma(g)=\lambda(g)$.

Finally, we consider some specific case.
\begin{enumerate}
\item For translation+parity+$SO(3)$, there are $2\times 2\times 1\times 2\times 1=8$ types of phases
\item For translation+parity+$D_2$, there are $2\times 2\times 4\times 2\times 4=128$ types of phases
\end{enumerate}

\subsection{Time reversal + On-site $G$}

Now consider a 1D spin system symmetric under both time reversal $T=v\otimes v ...\otimes v K$ and on-site unitary $u(g)\otimes u(g)...\otimes u(g)$. On each site, $T^2=vv^*=I$ and $u(g)$ forms a linear representation of $G$. Eq. \ref{eqn:U} and \ref{eqn:TR} are satisfied due to the two symmetries separately with some choice of $\alpha(g)$, $R(g)$, and $M$. $\alpha(g)$ labels the 1D representation the state forms under $G$, $R(g)\in \omega$ is the projective representation of $G$ on the boundary spin, and $\hat{T}=MK$ is the time reversal operator on the boundary spin which squares to $\beta(T) I=\pm I$. Note that while $\hat{T}^2=\pm I$ on the boundary spin, $T$ always square to $I$ on the physical spin at each site. 

Moreover, the two symmetries commute, i.e. $u(g)v=vu^*(g)$. This leads to non-trivial relations between $R(g)$ in Eq.\ref{eqn:U} and $M$ in Eq.\ref{eqn:TR}. In particular, suppose we act $G$ first followed by $T$, the matrices $A_{i}$ transform as,
\begin{equation}
\begin{array}{lll}
A_i \xrightarrow{G} A'_i &=& \sum_j u_{ij}(g)A_j = \alpha(g)R^{-1}(g)A_iR(g) \\
A'_i \xrightarrow{T} A''_i &=& \sum_j v_{ij} (A')^*_j \\
                                       &=& \alpha^*(g) (R^*)^{-1}(g)(\sum_j v_{ij} A^*_j)R^*(g) \\
                                       &=& \alpha^*(g) (R^*)^{-1}(g)M^{-1}A_iMR^*(g)
\end{array}
\end{equation}
Acting $T$ first followed by $G$ gives
\begin{equation}
\begin{array}{lll}
A_i \xrightarrow{T} A'_i &=& \sum_j v_{ij} A^*_j = M^{-1}A_iM \\
A'_i \xrightarrow{G} A''_i &=& \sum_j u_{ij}(g) A'_j \\
                                       &=& M^{-1}(\sum_j u_{ij}(g) A_j)M \\
                                       &=& \alpha(g) M^{-1}R^{-1}(g)A_i R(g)M
\end{array}
\end{equation}
Because $u(g)v=vu^*(g)$, the previous two transformations should be equivalent. That is
\begin{equation}
\begin{array}{l}
\alpha^*(g) (R^*)^{-1}(g)M^{-1}A_iMR^*(g) = \\
\alpha(g) M^{-1}R^{-1}(g)A_i R(g)M
\end{array}
\label{UT_equiv}
\end{equation}
Denote $Q=MR^*M^{-1}R^{-1}$. It follows that $A_i=\alpha^2(g)QA_iQ^{-1}$. Suppose that the MPS is injective with blocks larger than $n$ sites, hence $\mathcal{A}_I=A_{i_1}A_{i_2}...A_{i_n}$ satisfies $\mathcal{A}_I=\alpha^{2n}(g)Q\mathcal{A}_IQ^{-1}$. But $\mathcal{A}_I$ spans the whole space of $D\times D$ matrices, therefore $Q\propto I$. That is
\begin{equation}
M^{-1}R(g)M=e^{i\theta(g)}R^*(g)
\end{equation}
Moreover, it follows from Eq.\ref{UT_equiv} that $\alpha^2(g)=1$. That is, the 1D representation of $G$ must have order $2$. 

Similar to the parity + on-site $G$ case, we find, $\omega^2=1$, and $e^{i\theta(g)}$ form a 1D representation, denoted by $\gamma(g)$. Two $\gamma(g)$'s that differ by the square of a third 1D representation are equivalent.

Therefore, different phases with translation, 
time reversal, and on-site $G$ symmetries are labeled by 
\begin{enumerate}
\item $\beta(T)$, $\hat{T}^2=\pm I$ on the boundary spin
\item $\alpha(g)$, 1D representation of $G$, $\alpha^2(g)=1$.
\item $\omega$, projective representation of $G$ on boundary spin, $\omega^2=1$.
\item $\gamma(g) \in \mathcal{G}/\mathcal{G}_2$,1D representation of $G$ related to commutation relation between time reversal and on-site $G$, where $\mathcal{G}$ is the group of 1D representation of $G$, $\mathcal{G}_2$ is the group of 1D representation squared of $G$.
\end{enumerate}

Representative states are again given by Fig.\ref{fig:rep_2}. On-site white
pair forms 1D representation $\eta(g)$ for $G$ and is invariant under $T$.
Similar to the parity+on-site symmetry case, we can show that for any
$\omega$($\omega^2=1$) and 1D representation $\lambda(g)$ there exist
$D$-dimensional projective representation $R(g) \in \omega$ and matrix $M$ such
that $MM^*= I$ and $M^{-1}R(g)M=\lambda(g)R^*(g)$. Choose the inter-site black
pair to be composed of two $D$-dimensional spins. Define time reversal on this
pair to be $(M\otimes M)K$ and define on-site symmetry to be $R(g)\otimes
R(g)$. If the state of the black pair is chosen to be $(M\otimes I)
\sum_i|i\rangle\otimes |i\rangle$, $i=1,...,D$, it is easy to check that it is
invariant under time reversal and forms 1D representation $\lambda(g)$ for $G$
and contains projective representation $\omega$ at each end. Finally, define
the state of the inter-site grey pair to be $|0\>\otimes|1\>+\rho(T)|1\>\otimes|0\>$,
$\rho(T)=\pm 1$. Time reversal acts on it as
$(|0\>\<1|+\rho(T)|1\>\<0|)\otimes(|0\>\<1|+\rho(T)|1\>\<0|)K$. On-site $G$
acts trivially on it. The 1D spin state constructed as this is symmetric under
time reversal and on-site unitary $G$ and belongs to the SPT phase labeled by
$\beta(T)=\rho(T)$, $\alpha(g)=\eta(g)\lambda(g)$, $\omega$, and
$\gamma(g)=\lambda(g)$.

Applying the general classification result to specific cases we find
\begin{enumerate}
\item For Translation+$T+SO(3)$, there are $2\times 1\times 2\times 1=4$ types of phases
\item For Translation+$T+D_2$, there are $2\times 4\times 2\times 4=64$ types of phases
\end{enumerate}

If translation symmetry is not required, 
different phases with time reversal and on-site $G$ symmetries are labeled by 
\begin{enumerate}
\item $\beta(T)$, time reversal even/odd on the boundary spin
\item $\omega$, projective representation of $G$ on boundary spin, $\omega^2=1$.
\item $\gamma(g) \in \mathcal{G}/\mathcal{G}_2$,1D representation of $G$ related to commutation relation between time reversal and on-site $G$, where $\mathcal{G}$ is the group of 1D representation of $G$, $\mathcal{G}_2$ is the group of 1D representation squared of $G$.
\end{enumerate}
$\alpha(g)$ can no longer be used to distinguish
different phases. We find
\begin{enumerate}
\item For $T+SO(3)$, there are $2\times 2\times 1=4$ types of phases
\item For $T+D_2$, there are $2\times 2\times 4=16$ types of phases
\end{enumerate}

\subsection{Parity + Time reversal}

When parity is combined with time reversal, what SPT phases exist in 1D gapped spin systems? First we realize that due to parity and time reversal separately, different SPT phases exist labeled by different $\alpha(P)$, $\beta(P)$, $\beta(T)$ as defined in Eq.\ref{eqn:TR} and \ref{eqn:P}. $\alpha(P)$ labels parity even or odd, $\beta(P)$ labels parity even/odd between sites, and $\hat{T}^2=\beta(T) I$ on the boundary spin.

Does the commutation relation between parity and time reversal give more phases? The combined operation of parity first and time reversal next gives
\begin{equation}
\begin{array}{lll}
A_i \xrightarrow{P} A'_i &=& \sum_j w_{ij} A^T_j = \alpha(P)N^{-1}A_iN \\
A'_i \xrightarrow{T} A''_i &=& \sum_j v_{ij} (A')^*_j \\
                                       &=& \alpha(P) (N^{-1})^*(\sum_j v_{ij}A^*_j)N^* \\
                                       &=& \alpha(P) (N^{-1})^*M^{-1}A_iMN^*
\end{array}
\end{equation}
and the operation with time reversal first and parity next gives
\begin{equation}
\begin{array}{lll}
A_i \xrightarrow{T}  A'_i &=&  \sum_{j} v_{ij} A^*_j=M^{-1}A_iM \\
A'_i \xrightarrow{P} A''_i &=& \sum_j w_{ij} (A')^T_j \\
                                       &=& M^T(\sum_j w_{ij} A^T_j)(M^T)^{-1} \\
                                       &=& \alpha(P) M^TN^{-1}A_iN(M^T)^{-1}
\end{array}
\end{equation}
As parity and time reversal commute, $wv=vw^*$. The above
two operations should be equivalent. 
\begin{equation}
(N^{-1})^*M^{-1}A_iMN^* = M^TN^{-1}A_iN(M^T)^{-1}
\end{equation}
As $A_i$ is injective, $MN^*M^TN^{-1} \propto I$. That is
\begin{equation}
MN^{\dagger}MN^{\dagger} = e^{i\theta} I
\end{equation}
But $e^{i\theta}$ can be set to be $1$ by changing the phase of $M$ or $N$, therefore, the commutation relation does not lead to more distinct phases.

There are hence eight SPT phases with both parity and time reversal symmetry, labeled by 
\begin{enumerate}
\item $\alpha(P)$, parity even/odd;
\item $\beta(P)$, parity even/odd between sites;
\item $\beta(T)$, $\hat{T}^2=\pm I$ on boundary spins.
\end{enumerate} 
The representative states of each phase can be given as in Fig.\ref{fig:rep_2}. Each pair of spins forms a 1D representation of parity and time reversal. The on-site white pair is in the state $|0\>\otimes|1\> +\eta(P)|1\>\otimes|0\>$ with $\eta(P)=\pm 1$. Parity on this pair is defined as exchange of spins and time reversal as $K$. Therefore, this pair has parity $\eta(P)$ and is invariant under $T$. The inter-site black pair is in the state $|0\>\otimes|1\> +\lambda|1\>\otimes|0\>$ with $\lambda=\pm 1$. Parity acts on it as exchange of spins and time reversal as $(|0\>\<1|+\lambda|1\>\<0|)\otimes (|0\>\<1|+\lambda|1\>\<0|) K$. This pair therefore has parity $\lambda(P)=\lambda$ and is invariant under time reversal. Time reversal on one of the spins square to $\lambda(T) I=\lambda I$. Finally, the inter-site grey pair is in state $|0\>\otimes|1\> +\rho(P)|1\>\otimes|0\>$ with $\rho(P)=\pm 1$. Parity on this pair is defined as exchange of spins and time reversal as $K$. Therefore this pair has parity $\rho(P)$ and is invariant under $T$. Time reversal on one of the spins square to $I$. This state is in the SPT phase labeled by $\alpha(P)=\eta(P)\lambda(P)\rho(P)$, $\beta(P)=\lambda(P)\rho(P)$, $\beta(T)=\lambda(T)$.

\subsection{Parity + Time reversal + On-site $G$}

Finally we put parity, time reversal and on-site unitary symmetry together and ask how many SPT phases exist if the ground state does not break any of the symmetries. From Eq. \ref{eqn:U}, \ref{eqn:TR}, \ref{eqn:P}, we know that due to the three symmetries separately, states with different $\alpha(g)$, $\omega$, $\alpha(P)$, $\beta(P)$, $\beta(T)$ belong to different SPT phases. $\alpha(g)$ labels the 1D representation the state forms under $G$, $\omega$ is the projective representation of $G$ on the boundary spin, $\alpha(P)$ labels parity even or odd, $\beta(P)$ labels parity even/odd between sites, and $\hat{T}^2=\beta(T) I$ on the boundary spin.

Moreover, the commutation relation between parity, time reversal and on-site $G$ yields further conditions. 
The commutation relation between parity and on-site $G$ constrains $\omega^2=1$ and gives,
\begin{equation}
N^{-1}R(g)N=\gamma(g)R^*(g)
\end{equation}
$\gamma(g)$ is a 1D representation of $G$. $\gamma_1(g)$ and $\gamma_2(g)$ correspond to different SPT phases if and only if they are not related by the square of a third 1D representation.
The commutation relation between time reversal and on-site $G$ constrains $\alpha^2(g)=1$ and gives,
\begin{equation}
M^{-1}R(g)M=\gamma'(g)R^*(g)
\end{equation}
$\gamma'(g)$ is a 1D representation of $G$. $\gamma'_1(g)$ and $\gamma'_2(g)$ correspond to different SPT phases if and only if they are not related by the square of a third 1D representation.
The commutation relation between time reversal and parity gives,
\begin{equation}
MN^{\dagger}MN^{\dagger}\propto I
\end{equation}
which is equivalent to, because $N^T=\pm N$ and $M^T=\pm M$,
\begin{equation}
MN^*\propto NM^*
\end{equation}
Therefore, $MN^*$ and $NM^*$ conjugating $R(g)$ should give the same result
\begin{equation}
\begin{array}{lll}
(MN^*)R(g)(MN^*)^{-1} &=& \gamma(g)MR^*(g)M^{-1} \\
                      &=& \gamma(g)/\gamma'(g)R(g)
\end{array}
\end{equation}
On the other hand,
\begin{equation}
\begin{array}{lll}
(NM^*)R(g)(NM^*)^{-1} &=& \gamma'(g)NR^*(g)N^{-1} \\
                      &=& \gamma'(g)/\gamma(g)R(g)
\end{array}
\end{equation}
As $R(g)$ is nonzero, $\gamma'(g)=\pm \gamma(g)$.

$\gamma(g)$ and $\gamma'(g)$ are hence related by a 1D representation $\chi(g)$ which squares to $1$. As shown in Appendix \ref{append_C}, for fixed $\omega$, if $N$ and $R(g)$ exist that satisfy $N^{-1}R(g)N=\gamma(g)R^*(g)$, then any choice of $\gamma'(g)=\chi(g)\gamma(g)$($\chi^2(g)=1$) can be realized. The freedom in $\chi(g)$ is $\mathcal{G}/\mathcal{G}_2$.  Considering the degree of freedom in choosing $\gamma(g)$, the total freedom in $\{\gamma(g),\gamma'(g)\}$ is $(\mathcal{G}/\mathcal{G}_2)\times (\mathcal{G}/\mathcal{G}_2)$.

The SPT phases with parity, time reversal and on-site unitary symmetries are labeled by 
\begin{enumerate}
\item
$\alpha(g)$, 1D representation of $G$, $\alpha^2(g)=1$;
\item
$\omega$, projective representation of $G$ on boundary spin, $\omega^2=1$;
\item
$\alpha(P)$, parity even/odd;
\item
$\beta(P)$, parity even/odd between sites;
\item
$\beta(T)$, $\hat{T}^2=\pm I$ on the boundary spin;
\item
$\{\gamma(g),\gamma'{g}\} \in (\mathcal{G}/\mathcal{G}_2)\times (\mathcal{G}/\mathcal{G}_2)$, 1D representations of $G$, related to commutation relation between time reversal, parity and on-site $G$. 
\end{enumerate}

Representative states can be constructed as in Fig.\ref{fig:rep_2}.

Each pair is invariant(up to phase) under $G$, $T$ and $P$.

White pair: forms a 1D representation $\eta(g)$ of $G$, $\eta(P)$ of $P$ and is invariant under time reversal.

Black pair: $G$ acts non-trivially on it as $R(g)\otimes R(g)$. $R(g)$ is a $D$-dimensional projective representation and belongs to class $\omega$. According to appendices \ref{append_A}, \ref{append_B} and \ref{append_C}, for any 1D representation $\lambda(g)$ and order $2$ 1D representation $\chi(g)$, we can find matrices $N$ and $M$ such that 
$N^T= N$, $N^{-1}R(g)N = \lambda(g) R^*(g)$, $MM^*=I$, $M^{-1}R(g)M=\lambda'(g)R^*(g)=\chi(g)\lambda(g)R^*(g)$, $MN^*=NM^*$.
Now set the state of this pair to be $N\sum_i |i\rangle\otimes|i\rangle$, where $i=1...D$. Define parity as exchange of sites and time reversal as $(M\otimes M)K$. It can be checked that the state forms a 1D representation $\lambda(g)$ for $G$, has even parity and is invariant under time reversal. Time reversal squares to $I$ at each end.

Grey pair: $G$ acts trivially on it. The pair is in state $|0\>\otimes|1\> +\rho(P)|1\>\otimes|0\>$ with $\rho(P)=\pm 1$. Parity on this pair is defined as exchange of spins and time reversal as $(Y\otimes Y)^{(\rho(T)+1)/2}K$ with $\rho(T)=\pm 1$. Therefore this pair has parity $\rho(P)$ and is invariant under $T$. Time reversal on one of the spins square to $\rho(T)I$.

This state is representative of the SPT phase labeled by $\alpha(g)=\eta(g)\lambda(g)$, $\omega$, $\alpha(P)=\eta(P)\rho(P)$, $\beta(P)=\rho(P)$, $\beta(T)=\rho(T)$, $\gamma(g)=\lambda(g)$, $\gamma'(g)=\lambda'(g)$.

When $G=SO(3)$ or $G=D_2$, the classification result gives: 
\begin{enumerate}
\item For translation+$T+P+SO(3)$, there are 
$1\times 2\times 2\times 2\times 2 \times(1\times 1)=16$ types of phases
\item For translation+$T+P+D_2$, there are 
$4\times 2\times 2\times 2\times 2\times (4\times 4)=1024$ types of phases
\end{enumerate}

\begin{table}[tb]
 \centering
 \begin{tabular}{ |c|c| }
 \hline
 Symmetry of Hamiltonian   & Number of Different Phases \\ 
\hline
\hline
 None 		& 		1 		  \\ 
\hline
 $SO(3)$& 		$2$ 		  \\ 
\hline
 $D_2$& 		$2$ 	  \\ 
\hline
 $T$ & 	2 		  \\ 
\hline
 $SO(3)+ T$& 	4 		  \\ 
\hline
 $D_2+T$& 		16 	  \\ 
\hline
\hline
 Trans. +$U(1)$& 		$\infty$ 		  \\ 
\hline
 Trans. +$SO(3)$& 		$2$ 		  \\ 
\hline
 Trans. +$D_2$& 		$4\times 2=8$ 	  \\ 
\hline
\hline
 Trans. + $P$ & 	4 		  \\ 
\hline
 Trans. + $T$ & 	2 		  \\ 
\hline
 Trans. + $P+T$ & 	8 		  \\ 
\hline
\hline
 Trans. +$SO(3)+ P$& 	8 		  \\ 
\hline
 Trans. +$D_2+P$& 		128 	  \\ 
\hline
 Trans. +$SO(3)+ T$& 	4 		  \\ 
\hline
 Trans. +$D_2+T$& 		64 	  \\ 
\hline
 Trans. +$SO(3)+ P+T $& 	 16 		  \\ 
\hline
 Trans. +$D_2+P+T$ & 		1024 	  \\ 
\hline
 \end{tabular}
 \caption{Numbers of different 1D gapped quantum phases
that do not break any symmetry. $T$ stands for time reversal, $P$ stands for parity, and Trans. stands for translational symmetry.
}
 \label{table}
\end{table}

In table \ref{table}, we summarize the results obtained above and in
\Ref{CGW1107}.

\subsection{General classification for SPT phases}

Besides the cases discussed above, it is possible to have other types of symmetries in 1D spin systems. For example, there could be systems where time reversal and parity are not preserved individually but the combined action of them together defines a symmetry of the system. The general rule for classifying SPT phases under any symmetry is to classify all the projective representations of the total symmetry group, where on-site unitary symmetries should be represented with unitary matrices, on-site anti-unitary symmetries should be represented with anti-unitary matrices, and parity should be represented with anti-unitary matrices. Moreover, if translational symmetry is present, another independent label for SPT phases exists which corresponds to different 1D representations of the total symmetry group. In calculating this label, the representation of the total symmetry group is slightly different from the one for calculating projective representations. In particular,  parity should be represented unitarily, i.e. as a complex number, while on-site unitary/anti-unitary symmetries should still be represented unitarily/anti-unitarily.

\section{Classification with symmetry breaking}
\label{SB}

In \Ref{CGW1107} and previous sections we have only considered 1D gapped phases whose ground state does not break any symmetry and hence is non-degenerate. These SPT phases correspond to one section(labeled `SF' in Fig.\ref{fig:phase}) in the phase diagram for short range entangled states. Of course apart from SPT phases, there are symmetry breaking phases. It is also possible to have phases where the symmetry is only partly broken and the non-broken symmetry protects non-trivial quantum order. In this section, we combine symmetry breaking with symmetry protection and complete the classification for gapped phases in 1D spin systems. We find that 1D gapped phases are labeled by (1) the unbroken symmetry subgroup (2) SPT order under the unbroken subgroup. This result is the same as that in \Ref{SPC1032}.

\subsection{Matrix product representation of symmetry breaking}

Before we try to classify, we need to identify the class of systems and their gapped ground states that are under consideration. As we briefly discussed in the introduction, while the meaning of symmetry breaking is straight forward in classical system, this concept is more subtle in the quantum setting. A classical system is in a symmetry breaking phase if each possible ground state has lower symmetry than the total system. For example, the classical Ising model has a spin flip symmetry between spin up $|\uparrow\>$ and spin down $|\downarrow\>$ which neither of its ground states $|\uparrow\uparrow...\uparrow\>$ and $|\downarrow\downarrow...\downarrow\>$ have. However, in quantum Ising model $H=\sum_{<i,j>} -\sigma^i_z\sigma^j_z$, the ground space contains not only these two states, but also any superposition of them, including the state $|\uparrow\uparrow...\uparrow\> +|\downarrow\downarrow...\downarrow\>$ which is symmetric under spin flip. This state is called the `cat' state or the GHZ state in quantum information literature. In fact, if we move away from the exactly solvable point by adding symmetry preserving perturbations(such as transverse field $B_x\sum_i \sigma^i_x$) and solve for the ground state at finite system size, we will always get a state symmetric under spin flip. Only in the thermodynamic limit does the ground space become two dimensional. How do we tell then whether the ground states of the system spontaneously break the symmetry?

With matrix product representation, the symmetry breaking pattern can be easily seen from the matrices.\cite{FNW9243,PVW0701,N9665} Suppose that we solved a system with certain symmetry at finite size and found a unique minimum energy state which has the same symmetry. To see whether the system is in symmetry breaking phase, we can write this minimum energy state in matrix product representation. The matrices in the representation can be put into a `canonical' form \cite{PVW0701} which is block diagonal
\begin{equation}
A_i = \begin{bmatrix} A^{(0)}_i & & \\ & A^{(1)}_i  & \\ & & \ddots \end{bmatrix}
\label{canon_form}
\end{equation}
where the double tensor for each block $\mathbb{E}^{(k)} =\sum_i A_i^{(k)} \otimes (A_i^{(k)})^*$ has a non-degenerate largest eigenvalue $\lambda_i$. If in the thermodynamic limit, the canonical form contains only one block, this minimum energy state is short range correlated and the system is in a symmetric phase as discussed in \Ref{CGW1107} and the previous section. However, if the canonical form splits into more than one block with equal largest eigenvalue(set to be $1$) when system size goes to infinity, then we say the symmetry of the system is spontaneously broken in the ground states.

The symmetry breaking interpretation of block diagonalization of the canonical form can be understood as follows. Each block of the canonical form $A^{(k)}_i$ represents a short range correlated state $|\psi_k\>$. Note that here by correlation we always mean connected correlation $<O_1O_2>-<O_1><O_2>$. Therefore, the symmetry breaking states $|\uparrow\uparrow...\uparrow\>$ and $|\downarrow\downarrow...\downarrow\>$ both have short range correlation. Two different short range correlated states $|\psi_k\>$ and $|\psi_{k'}\>$ have zero overlap $\<\psi_{k'}|\psi_k\>=0$ and any local observable has zero matrix element between them $\<\psi_{k'}|O|\psi_k\>=0$. The ground state represented by $A_i$ is an equal weight superposition of them $|\psi\>=\sum_k |\psi_k\>$. Actually the totally mixed state $\rho=\sum_k |\psi_k\>\<\psi_k|$ has the same energy as $|\psi\>$ as $\<\psi_{k'}|H|\psi_k\>=0$ for $k' \neq k$. Therefore, the ground space is spanned by all $|\psi_k\>$'s. Consider the operation which permutes $|\psi_k\>$'s. This operation keeps ground space invariant and can be a symmetry of the system. However, each short range correlated ground state is changed under this operation. Therefore, we say that the ground states spontaneously break the symmetry of the system. 

This interpretation allows us to study symmetry breaking in 1D gapped systems by studying the block diagonalized canonical form of matrix product states. Actually, it has been shown that for any such state a gapped Hamiltonian can be constructed having the space spanned by all $|\psi_k\>$'s as ground space\cite{N9665}. Therefore, we will focus on finite dimensional matrix product states in block diagonal canonical form for our classification of gapped phases involving symmetry breaking.

\subsection{Classification with combination of symmetry breaking and symmetry fractionalization}

We will consider class of systems with certain symmetry and classify possible phases. For simplicity of notation, we will focus on on-site unitary symmetry. With slight modification, our results also apply to parity and time reversal symmetry and their combination. Suppose that the system has on-site symmetry of group $G$ which acts as $u(g)\otimes u(g)...\otimes u(g)$. It is possible that this symmetry is not broken, totally broken or partly broken in the ground state. In general, suppose that there is a short range correlated ground state $|\psi_0\>$ that is invariant under only a subgroup $G'$ of $G$. Of course, different $G'$'s represent different symmetry breaking patterns and hence lead to different phases. Moreover, $|\psi_0\>$ could have different symmetry protected order under $G'$ which also leads to different phases. In the following we are going to show that these two sets of data: (1) the invariant subgroup $G'$ and (2) the SPT order under $G'$ describe all possible 1D gapped phases. Specifically we are going to show that if two systems symmetric under $G$ have short range correlated ground states $|\psi_0\>$ and $|\t{\psi}_0\>$ which are invariant under the same subgroup $G'$ and $|\psi_0\>$ and $|\t{\psi}_0\>$ have the same symmetry protected order under $G'$, then the two systems are in the same phase. We are going to construct explicitly a path connecting the ground space of the first system to that of the second system without closing gap and breaking the symmetry of the system.

Assume that $|\psi_0\>$ has a $D$-dimensional MPS representation $A^{(0)}_i$ which satisfies
\begin{equation}
\sum_{ij} u(g')_{ij} A^{(0)}_j = \alpha(g')M^{-1}(g') A^{(0)}_i M(g')
\end{equation}
where $g'\in G'$, $\alpha(g')$ is a 1D representation of $G'$ and $M(g')$ form a projective representation of $G'$.

Suppose that $\{h_0,h_1,...,h_m\}$ are representatives from each of the cosets of $G'$ in $G$, $h_0\in G'$. Define $|\psi_k\>=u(h_k)\otimes u(h_k) \otimes ...\otimes u(h_k)|\psi_0\>$, $k=1,...,m$. $|\psi_k\>$ are hence each short-range correlated and orthogonal to each other. The ground space of the system is spanned by $|\psi_k\>$. The MPS representation of $|\psi_k\>$ is then $A^{(k)}_i=\sum_j u(h_k)_{ij} A^{(0)}_j$, which satisfies similar symmetry conditions
\begin{equation}
\sum_{ij} u(h_kg'h_k^{-1})_{ij} A^{(k)}_j = \alpha(g')M^{-1}(g') A^{(k)}_i M(g')
\end{equation}

To represent the whole ground space, put all $A^{(k)}_i$ into a block diagonal form and define
\begin{equation}
A_i = \begin{bmatrix} A^{(0)}_i & & \\ & \ddots  & \\ & & A^{(m)}_i \end{bmatrix}
\end{equation}
The state represented by $A_i$ is then the superposition of all $|\psi_k\>$, which is equivalent to the maximally mixed ground state with respect to any local observable.

Under any symmetry operation $g\in G$, $A_i$ changes as
\begin{equation}
\sum_j u(g)_{ij} A_j = P(g) \Theta(g)Q(g)A_iQ^{-1}(g)P^{-1}(g)
\label{A_symm}
\end{equation}
where
\begin{equation}
\Theta(g) = \begin{bmatrix} \alpha{(g'_{g,0})} & &  \\ & \ddots & \\ & & \alpha{(g'_{g,m})}  \end{bmatrix} \otimes I_n
\end{equation}
\begin{equation}
P(g) = p(g) \otimes I_n
\end{equation}
with $p(g)$ $m\times m$ permutation matrices and form a  linear representation of $G$.
\begin{equation}
Q(g) = \begin{bmatrix} M(g'_{g,0}) &  & \\ & \ddots & \\ & & M(g'_{g,m}) \end{bmatrix}
\label{Q(g)}
\end{equation}

To classify phases, we first deform the state into a simpler form by using the double tensor.

Define the double tensor for the whole state as
\begin{equation}
\mathbb{E} = \sum_i A_i\otimes A_i^*
\end{equation}
As $A_i=\oplus_k  A_i^{(k)}$
\begin{equation}
\mathbb{E} = (\oplus_k \mathbb{E}^{(k)}) \oplus (\oplus_{k \neq k'} \mathbb{E}^{(kk')})
\end{equation}
where $\mathbb{E}^{(k)} =\sum_i A_i^{(k)} \otimes (A_i^{(k)})^*$, $\mathbb{E}^{(kk')} = \sum_i A_i^{(k)} \otimes (A_i^{(k')})^*$.  As $A_i^{(k)}$ for different $k$ only differ by a local unitary on the physical index $i$, $\mathbb{E}^{(k)}$ all have the same form with a single non-degenerate largest eigenvalue. WLOG, we set it to be $1$ and denote the corresponding eigensector as $\mathbb{E}^{(k)}_0$. On the other hand, $\<\psi_k'|\psi_k\>  = \lim_{n\to \infty} Tr \left(\mathbb{E}^{(kk')}\right)^n =0$, therefore, $\mathbb{E}^{(kk')}$ all have eigenvalues strictly less than $1$. Define
\begin{equation}
\mathbb{E}_0 = \oplus_k \mathbb{E}^{(k)}_0
\end{equation}
$\mathbb{E}_0$ is the eigenvalue $1$ sector of $\mathbb{E}$ and $\mathbb{E}_1=\mathbb{E}-\mathbb{E}_0$ has eigenvalues strictly less than $1$.

The symmetry condition on $A_i$ can be translated to $\mathbb{E}$ as
\begin{equation}
\mathbb{E} = \bar{P}(g) \bar{\Theta}(g) \bar{Q}(g) \mathbb{E} \bar{Q}^{-1}(g) \bar{P}^{-1}(g)
\label{E_symm1}
\end{equation}
where $\bar{P}(g)=P(g)\otimes P(g)$($P(g)$ is real), $\bar{\Theta}(g)=\Theta(g)\otimes \Theta^*(g)$, $\bar{Q}(g)=Q(g)\otimes Q^*(g)$. 

Matching the eigenvalue $1$ sector on the two side of Eq. \ref{E_symm1}, it is clear to see that
\begin{equation}
\mathbb{E}_0 = \bar{P}(g) \bar{\Theta}(g) \bar{Q}(g)  \mathbb{E}_0 \bar{Q}^{-1}(g) \bar{P}^{-1}(g)
\end{equation}
It follows that,
\begin{equation}
\mathbb{E}_	1 = \bar{P}(g) \bar{\Theta}(g) \bar{Q}(g)  \mathbb{E}_1 \bar{Q}^{-1}(g) \bar{P}^{-1}(g)
\end{equation}
That is, $\mathbb{E}_0$ and $\mathbb{E}_1$ satisfy the symmetry condition separately.

Now define the deformation path of the double tensor analogous to \Ref{CGW1107} as
\begin{equation}
\mathbb{E}(t) = \mathbb{E}_0 + \left(1-\frac{t}{T}\right)\mathbb{E}_1
\end{equation}
We will show that as $t$ increases from $0$ to $T$, this corresponds to a deformation of the ground space to a fixed point form while the system remains gapped and symmetric under $G$.

First, it can be checked that for $0\leq t \leq T$, $\mathbb{E}(t)$ remains a valid double tensor and satisfies symmetry condition Eq.\ref{E_symm1}. Decomposing $\mathbb{E}(t)$ back into matrices, $A_i(t)$ necessarily contains $m$ blocks each with finite correlation length. The fact that the total state remains gapped is proven by \Ref{N9665}. Because two equivalent double tensor can only differ by a unitary transformation on the physical index, the symmetry condition Eq.\ref{E_symm1} for $\mathbb{E}(t)$ gives that there exist unitary transformations $u(g)(t)$ such that $A_i(t)$ transform in the same way as $A_i$ in Eq.\ref{A_symm}. The symmetry operation can be defined continuously for all $t$.

At $t=T$, the state is brought to the fixed point form $\mathbb{E}(T) = \mathbb{E}_0 = \sum_k \mathbb{E}^{(k)}_0$. Each block $k$ represents a dimer state as in Fig.\ref{fig:rep_1} with entangled pairs between neighboring sites supported on dimension $1_k,...,D_k$, $|EP_k\> = \lambda_{i_k} |i_ki_k\>$.  For different blocks $k$ and $k'$, $|i_k\> \perp |i_{k'}\>$. The total Hilbert space on one site is $(D\times m)^2$ dimensional. The symmetry operation on one site can then be defined as $u(g)=P(g)Q^*(g)\Theta(g)\otimes P(g)Q(g)$. With continuous deformation that does not close gap or violate symmetry, every gapped ground space is mapped to such a fixed point form. If we can show that two fixed point state with the same unbroken symmetry $G'$ and SPT order under $G'$ are in the same phase, we can complete the classification for combined symmetry breaking and SPT order.

Suppose that $|\psi_0\>$ and $|\t{\psi}_0\>$ are short range correlated fixed point ground states of two systems symmetric under $G$. Symmetry operations are defined as $u(g)$ and $\t u(g)$ respectively. $|\psi_0\>$ and $|\t{\psi}_0\>$ are symmetric under the same subgroup $G'$ and has the same SPT order. As shown in \Ref{CGW1107}, $|\psi_0\>$ and $|\t{\psi}_0\>$ can be mapped to each other with local unitary transformation $W_0$ that preserves $G'$ symmetry. The other short range correlated ground states can be obtained as $|\psi_k\>=u(h_k)|\psi_0\>$ and $|\t{\psi}_k\>=\t u(h_k)|\t{\psi}_0\>$. At fixed point, $|\psi_k\>$($|\t{\psi}_k\>$) are supported on orthogonal dimensions for different $k$ and $u(h_k)$ and $\t u(h_k)$ maps between these support spaces. Therefore, we can consistently define local unitary operations mapping between $|\psi_k\>$ and $|\t{\psi}_k\>$ as $W_k=\t u(h_k)W_0u^{\dagger}(h_k)$ and the total operation is $W=\oplus_k W_k$. $W$ as defined is a local unitary transformation symmetric under $G$ that maps between two fixed point gapped ground states with the same unbroken symmetry $G'$ and SPT order under $G'$. Combined with the mapping from a general state to its fixed point form, this completes our proof that 1D gapped phases are labeled by unbroken symmetry $G'$ and SPT order under $G'$.

\section{Application: 1D fermion SPT phases}
\label{fermion}

Although our previous discussions have been focused on spin systems, it actually also applies to fermion systems. Because in 1D, fermion systems and spin systems can be mapped to each other through Jordan Wigner transformation, we can classify fermionic phases by classifying corresponding spin phases. Specifically, for a class of fermion systems with certain symmetry we are going to 1. identify the corresponding class of spin systems by mapping the symmetry to spin 2. classify possible spin phases with this symmetry, including symmetry breaking and symmetry fractionalization 3. map the spin phases back to fermions and identify the fermionic order. In the following we are going to apply this strategy to 1D fermion systems in the following four cases respectively: no symmetry(other than fermion parity), time reversal symmetry for spinless fermions, time reversal symmetry for spin half integer fermions, and $U(1)$ symmetry for fermion number conservation. Our classification result is consistent with previous studies in \Ref{LK1103,TPB1102}. One special property of fermionic system is that it always has a fermionic parity symmetry. That is, the Hamiltonian is a sum of terms composed of even number of creation and annihilation operators. Therefore, the corresponding spin systems we classify always have an on-site $Z_2$ symmetry. Note that this approach can only be applied to systems defined on an open chain. For system with translation symmetry and periodic boundary condition, Jordan Wigner transformation could lead to non-local interactions in the spin system. 

\subsection{Fermion Parity Symmetry Only}

For a 1D fermion system with only fermion parity symmetry, how many gapped phases exist? 

To answer this question, first we do a Jordan-Wigner transformation and map the fermion system to a spin chain. The fermion parity operator $P_f = \prod (1-2a^{\dagger}_ia_i)$ is mapped to an on-site $Z_2$ operation. On the other hand, any 1D spin system with an on-site $Z_2$ symmetry can always be mapped back to a fermion system with fermion parity symmetry(expansion of local Hilbert space maybe necessary). As the spin Hamiltonian commute with the $Z_2$ symmetry, it can be mapped back to a proper physical fermion Hamiltonian. Therefore, the problem of classifying fermion chains with fermion parity is equivalent to the problem of classifying spin chains with $Z_2$ symmetry.

There are two possibilities in spin chains with $Z_2$ symmetry: (1) the ground state is symmetric under $Z_2$. As $Z_2$ does not have non-trivial projective representation, there is one symmetric phase. (If translational symmetry is required, systems with even number of fermions per site are in a different phase from those with odd number of fermions per site. This difference is somewhat trivial and we will ignore it.) (2) the ground state breaks the $Z_2$ symmetry. The ground state will be two-fold degenerate. Each short-range correlated ground state has no particular symmetry($G'=I$) and they are mapped to each other by the $Z_2$ operation. There is one such symmetry breaking phases. These are the two different phases in spin chains with $Z_2$ symmetry.

This tells us that there are two different phases in fermion chains with only fermion parity symmetry. But what are they? First of all, fermion states cannot break the fermion parity symmetry. All fermion states must have a well-defined parity. Does the spin symmetry breaking phase correspond to a real fermion phase?

The answer is yes and actually the spin symmetry breaking phase corresponds to a $Z_2$ symmetric fermion phase. Suppose that the spin system has two short-range correlated ground states $|\psi_0\>$ and $|\psi_1\>$. All connected correlations between spin operators decay exponentially on these two states. Mapped to fermion systems, $|\psi^f_0\>$ and $|\psi^f_1\>$ are not legitimate states but $|\t{\psi}^f_0\>=|\psi^f_0\> + |\psi^f_1\>$ and $|\t{\psi}^f_1\>=|\psi^f_0\> - |\psi^f_1\>$ are. They have even/odd parity respectively. In spin system, $|\t{\psi}_0\>$ and $|\t{\psi}_1\>$ are not short range correlated states but mapped to fermion system they are. To see this, note that any correlator between bosonic operators on the $|\t{\psi}^f_0\>$ and $|\t{\psi}^f_1\>$are the same as that on $|\psi_0\>$ and $|\psi_1\>$ and hence decay exponentially. Any correlator between fermionic operators on the $|\t{\psi}^f_0\>$ and $|\t{\psi}^f_1\>$ gets mapped to a string operator on the spin state, for example $a^{\dagger}_ia_j$ is mapped to $(X-iY))_iZ_{i+1}...Z_{j-1}(X-iY)_j$, which also decays with separation between $i$ and $j$. Therefore, the symmetry breaking phase in spin chain corresponds to a fermionic phase with symmetric short range correlated ground states. 
The degeneracy can be understood as isolated Majorana modes at the two ends of the chain\cite{K0131,K0922}.

On the other hand, the short-range correlated ground state in spin symmetric phase still correspond to short-range correlated fermion state after JW transformation. Therefore, the symmetric and symmetry breaking phases in spin system both correspond to symmetric phases in fermion system. The two fermion phases cannot be connected under any physical fermion perturbation.
 
\subsection{Fermion Parity and $T^2=1$ Time Reversal}

Now consider the more complicated situation where aside from fermion parity, there is also a time reversal symmetry $\mathcal{T}$. $\mathcal{T}$ acts as an anti-unitary $T=UK$ on each site. In this section we consider the case where $T^2=1$(spinless fermion). 

So now the total symmetry for the fermion system is the $Z_2$ fermion parity symmetry $P_f$ and $T^2=1$ time reversal symmetry. $T$ commutes with $P_f$. The on-site symmetry group is a $Z_2\times Z_2$ group and has four elements $G=\{I,T,P_f,TP_f\}$. Mapped to spin system, the symmetry group structure is kept. 

The possible gapped phases for a spin system with on-site symmetry $G=\{I,T,P_f,TP_f\}$ include: (1) $G'=G$. Following discussion in section \ref{CS} we find that it has four different projective representations. Examples of the four representations are a.$\{I, K, Z, KZ\}$, b. $\{I,iYK, Z, iYKZ\}$, c. $\{I,iYKZ\otimes I, I\otimes Z, iYKZ\otimes Z\}$ d. $\{I, K, Y, KY\}$. There are hence four different symmetric phases. (If translational symmetry is required, the number is multiplied by $2$ due to $\alpha(Z_2)$) (2) $G'=\{I,P_f\}$ with no non-trivial projective representation, the time reversal symmetry is broken. There is one such phase.  (If translational symmetry is required, there are two phases) (3) $G'=\{I,T\}$, with two different projective representations(time reversal squares to $\pm I$ on boundary spin). The $Z_2$ fermion parity is broken. There are two phases in this case.  (4) $G'=\{I,TP_f\}$, with two different projective representations. The fermion parity symmetry is again broken. Two different phases. (5) $G'=I$, no projective representation, all symmetries are broken.

Mapped back to fermion systems, fermion parity symmetry is never broken. Instead, the $P_f$ symmetry breaking spin phases are mapped to fermion phases with Majorana boundary mode on the edge as discussed in the previous section. Therefore the above spin phases correspond in the fermion system to: (1) Four different symmetric phases (2) One time reversal symmetry breaking phase. (3) Two symmetric phases with Majorana boundary mode (4) Another two symmetric phases with Majorana boundary mode. (5) One time reversal symmetry breaking phase. (1)(3)(4) contains the eight symmetric phases for time reversal invariant fermion chain with $T^2=1$. This is consistent with previous studies in \Ref{LK1009,TPB1102}.

\subsection{Fermion Parity and $T^2 \neq I$ Time Reversal}

When $T^2 \neq I$, the situation is different. This happens when we take the fermion spin into consideration and for a single particle, time reversal is defined as $e^{i\pi\sigma_y}K$. With half integer spin, $\left(e^{i\pi\sigma_y}K\right)^2=-I$. Note that for every particle the square of time reversal is $-I$, however when we write the system in second quantization as creation and annihilation operator on each site, the time reversal operation defined on each site satisfies  $T^2=P_f$. Therefore, the symmetry group on each site is a $Z_4$ group $G=\{I,T,P_f,TP_f\}$. To classify possible phases, we first map everything to spin.

The corresponding spin system has on-site symmetry $G=\{I,T,P_f,TP_f\}$. $T^2=P_f$, $P_f^2=I$. The possible phases are: (1) $G'=G$, with two possible projective representations, one with $T^4=I$, the other with $T^4=-I$. Example for the latter includes $T=(1/\sqrt{2})(X+Y)K$. Therefore, there are two possible symmetric phases. (If translational symmetry is required, there are four phases.) (2) $G'=\{I,P_f\}$, the time reversal symmetry is broken. One phase. (If translational symmetry is required, there are two phases.) (3) $G'=I$, all symmetries are broken. One phase.

Therefore, the fermion system has the following phases: (1) Two symmetric phases (2) One time reversal symmetry breaking phase. (3) One time reversal symmetry breaking phase with Majorana boundary mode. (1) contains the time reversal symmetry protected topological phase. Models in this phase can be constructed by first writing out the spin model in the corresponding spin phase and then mapping it to fermion system with Jordan-Wigner transformation.

\subsection{Fermion Number Conservation}

Consider the case of a gapped fermion system with fixed fermion number. This
corresponds to an on-site $U(1)$ symmetry, $e^{i\theta N}$. Mapped to spins,
the spin chain will have an on-site $U(1)$ symmetry. This symmetry cannot be
broken and $U(1)$ does not have a non-trivial projective representation. One
thing special about $U(1)$ symmetry though, is that it has an infinite family
of 1D representations. If translational symmetry is required, fermion number per
site is a good quantum number and labels different phases. Therefore, mapped back to fermions,
there is an infinite number of phases with different average number of fermions per site.

\section{Conclusion}
\label{conclude}

In this paper, we complete the classification of gapped phases in 1D spin
systems with various symmetries. Based on our classification of symmetry
protected topological phases  with on-site unitary, parity or time reversal
symmetry in \Ref{CGW1107}, we give explicit results in this paper for the classification of SPT
phases with combined on-site unitary, parity and/or time reversal symmetry. A general rule is also given for the classification of SPT phases with any symmetry group. Moreover, we considered the classification of phases with possible (partial) symmetry breaking. We find that
1D gapped spin phases with symmetry of group $G$ are basically labeled
by (1) the unbroken symmetry subgroup $G'$, (2) projective representations of $G'$. Note that in calculating projective representations of $G'$, on-site unitary symmetries are represented unitarily while parity and on-site anti-unitary symmetries are represented anti-unitarily. We apply this classification result to interacting 1D fermion
systems, which can be mapped to spin systems with Jordan-Wigner transformation,
and classify possible gapped phases with no symmetry, time reversal symmetry
and also fermion number conservation.

We would like to thank Andreas Ludwig, and Zheng-Xin Liu for very helpful
discussions.  This research is supported by  NSF Grant No. DMR-1005541.

\appendix

\section{Existence of $N$ such that $N^{-1}R(g)N=R^*(g)$}
\label{append_A}

In this section we will show that for any class $\omega$ of projective representation of group $G$ which satisfies $\omega^2=1$, there is a projective representation $R(g) \in \omega$ and a symmetric matrix $N$($N^T=N$) such that $N^{-1}R(g)N=R^*(g)$. 

Suppose that $R_0(g)$ is a $d$-dimensional projective representation in class $\omega$. Because $\omega^2=1$, $R^*_0(g)$ is also a projective representation in this class. And so is
\begin{equation}
R(g)=\begin{bmatrix} R_0(g) & \\  & R^*_0(g) \end{bmatrix}
\end{equation}
Define
\begin{equation}
N=\begin{bmatrix}  & I \\  I &  \end{bmatrix}
\end{equation}
Where $I$ is a $d$-dimensional identity matrix. It can be checked that $N^T=N$ and $N^{-1}R(g)N=R^*(g)$.
$\Box$

\section{Freedom in commutation relation between parity/time reversal and on-site $G$}
\label{append_B}

We will show in this section that for a fixed factor system $\omega$($\omega^2=1$), if there exists projective representation $R(g)$ and symmetric or antisymmetric matrix $N$($N^T=\pm N$), such that $N^{-1}R(g)N=\gamma(g)R^*(g)$ for one 1D representation $\gamma(g)$, then there are other $R'(g)$ and $N'$ which satisfy the relation for any other $\gamma'(g)$.

Suppose that $\gamma'(g)=\alpha(g)\gamma(g)$.

Define
\begin{equation}
R'(g)=\begin{bmatrix} R(g) & \\ & \alpha(g)R(g) \end{bmatrix}, N'=\begin{bmatrix} N&  \\ & N\end{bmatrix}\begin{bmatrix} & I \\I & \end{bmatrix}
\end{equation}

$R'(g)$ is another projective representation with factor system $\omega$ and ${N'}^T=\pm N'$.

Moreover it can be checked that
\begin{equation}
{N'}^{-1}R'(g)N=\alpha(g)\gamma(g)R'=\gamma'(g)R'
\end{equation}

$\Box$

\section{Freedom in commutation relation between on-site $G$, time reversal and parity}
\label{append_C}

We will show in this section that for a fixed factor system $\omega$($\omega^2=1$), if there exists projective representation $R(g)$ and symmetric or antisymmetric matrix $N$($N^T=\pm N$), such that $N^{-1}R(g)N=\gamma(g)R^*(g)$ for one 1D representation $\gamma(g)$, Then there exist $R'(g)$, $N'$(${N'}^T=\pm N'$), $M'$(${M'}^T=\pm M'$), such that $M'{N'}^*=N'{M'}^*$, ${N'}^{-1}R'(g)N'=\gamma(g){R'}^*(g)$, ${M'}^{-1}R'(g)M'=\gamma'(g){R'}^*(g)$, for any $\gamma'(g)=\chi(g)\gamma(g)$, $\chi^2(g)=1$.

Define
\begin{equation}
R'(g)=\begin{bmatrix} R(g) &   \\ & \chi(g)R(g)  \end{bmatrix}
\end{equation}
\begin{equation}
N'=\begin{bmatrix} N &   \\ & N  \end{bmatrix}
\end{equation}
\begin{equation}
M'=\begin{bmatrix}  & N  \\ N &  \end{bmatrix}
\end{equation}

It can be checked that, $R'(g)$ is a projective representation with factor system $\omega$. ${N'}^T=\pm N'$, and ${M'}^T=\pm M'$. Moreover,
\begin{equation}
M'{N'}^*=\begin{bmatrix}   & NN^*  \\ NN^* &  \end{bmatrix}=N'{M'}^*
\end{equation}
\begin{equation}
{N'}^{-1}R'(g)N' =  \gamma(g){R'}^*(g)
\end{equation}
\begin{equation}
{M'}^{-1}R'(g)M' = \chi(g)\gamma(g){R'}^*(g) =  \gamma'(g){R'}^*(g)
\end{equation}

$\Box$

\end{document}